\documentclass[footinbib,reprint, amsmath,amssymb,showpacs,floatfix, aps, prl, twocolumn, superscriptaddress]{revtex4-1}
\usepackage{graphicx}
\usepackage{dcolumn}
\usepackage{bm}
\usepackage{color}
\usepackage{hyperref}
\hypersetup{colorlinks=true,citecolor=blue,linkcolor=blue, urlcolor=blue}
\newcolumntype{M}[1]{>{\centering\arraybackslash}m{#1}}
\newcolumntype{N}{@{}m{0pt}@{}}

\def\nn{\nonumber}
\def\({\left(}
\def\){\right)}
\def\[{\left[}
\def\]{\right]}

\def\d{\delta}
\def\e{\epsilon}

\def\w{\omega}
\def\v{\nu}

\def\bfQ{\mathbf{Q}}

\def\bfk{\mathbf{k}}

\def\bfq{\mathbf{q}}

\def\>{\rangle}
\def\<{\langle}

\begin{document}
\title{Exciton-Phonon Interaction and Relaxation Times from First Principles}

\author{Hsiao-Yi Chen}

\affiliation {Department of Applied Physics and Materials Science, 
California Institute of Technology, Pasadena, California 91125}
%\affiliation {Department of Physics, 
%California Institute of Technology, Pasadena, California 91125}
\author{ Davide Sangalli}
\affiliation {CNR-ISM, Division of Ultrafast Processes in Materials (FLASHit), Area della Ricerca di Roma 1, Monterotondo Scalo, Italy}
\affiliation {Dipartimento di Fisica, Università degli Studi di Milano, via Celoria 16, I-20133 Milano, Italy}

\author{Marco Bernardi}
\email{bmarco@caltech.edu}
\affiliation {Department of Applied Physics and Materials Science, 
California Institute of Technology, Pasadena, California 91125}

\begin{abstract}
\noindent
Electron-phonon ($e$-ph) interactions are key to understanding the dynamics of electrons in materials, and can be modeled accurately from first-principles. 
However, when electrons and holes form Coulomb-bound states (excitons), quantifying their interactions and scattering processes with phonons remains an open challenge. Here we show a rigorous approach for computing exciton-phonon (ex-ph) interactions and the associated exciton dynamical processes from first principles. Starting from the \textit{ab initio} Bethe-Salpeter equation, we derive expressions for the ex-ph matrix elements and relaxation times. 
We apply our method to bulk hexagonal boron nitride, for which we map the ex-ph relaxation times as a function of exciton momentum and energy, analyze the temperature and phonon-mode dependence of the ex-ph scattering processes, and accurately predict the phonon-assisted photoluminescence. The approach introduced in this work is general and provides a framework for investigating exciton dynamics in a wide range of materials. %from first principles 
\end{abstract}
%\pacs{--}
\maketitle
\begin{figure}
\includegraphics[scale=0.23]{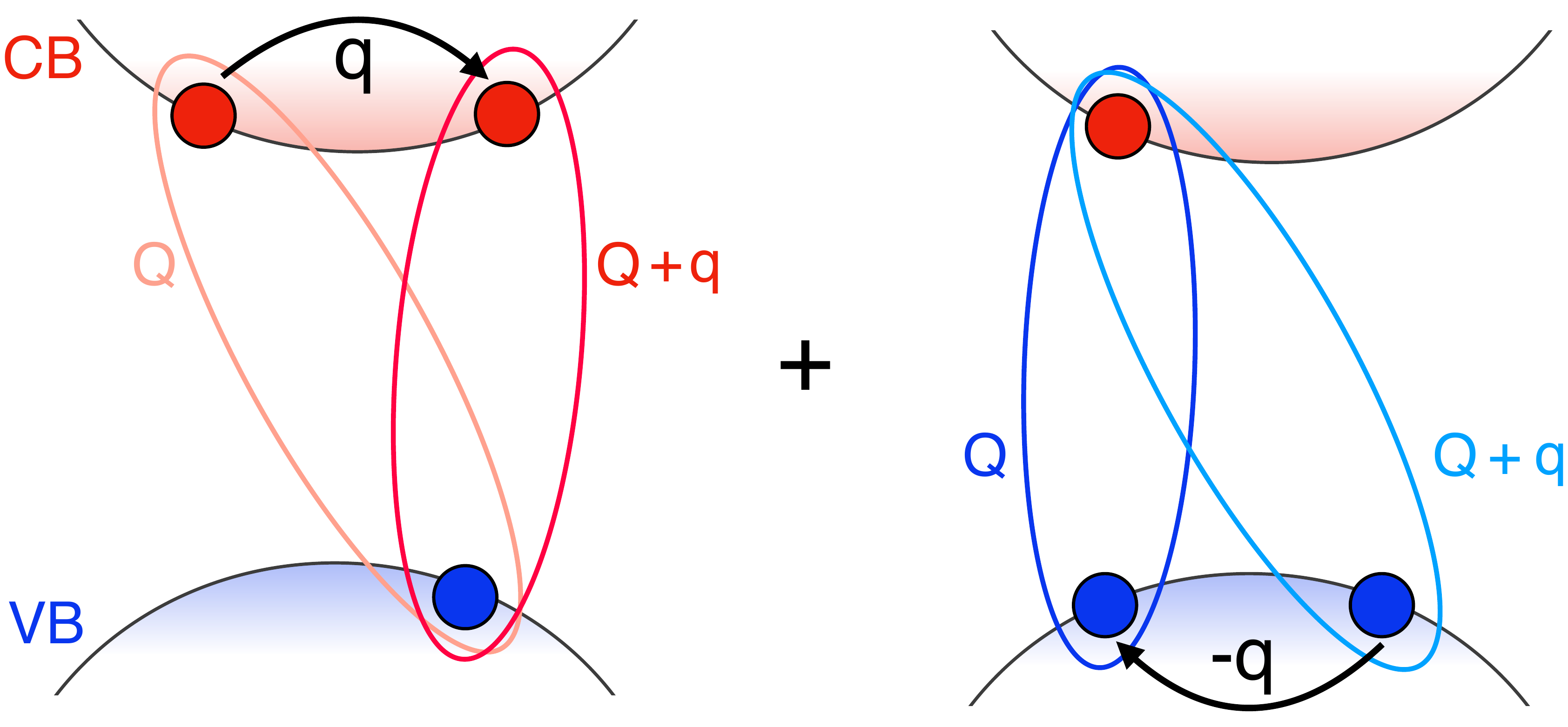}
\caption{Schematic of the exciton-phonon interaction in Eq.~(\ref{Eq:exph_coupling}), which can be viewed as a superposition of electron-phonon and hole-phonon scattering events, weighted by the wave functions of the initial and final exciton states. }
\label{Fig:scattering}
\end{figure}

 
% INTRODUCTION
Excitons are electron-hole pairs bound by the Coulomb interaction, and have been at the center of solid-state research for decades~\cite{Frenkel, knox1963, kohn1967}. 
They are essential for optoelectronic~\cite{quantum-wells, scholes2011excitons, bernardi2013extraordinary} and quantum technologies~\cite{Sham-2003, dark-exciton-2010, Xu2019}, 
and are actively investigated in materials ranging from quantum dots~\cite{gammon-qdots-2002} to two-dimensional semiconductors~\cite{Heinz-review,yuan2017exciton}, organic crystals~\cite{exciton-organic}   
and oxides~\cite{oxide-rydberg}. 
Exciton dynamics is probed with ultrafast optical or device measurements~\cite{marie20152d,steinleitner2017direct,oliver2018recent}; theories  
that can shed light on microscopic exciton processes and assist experiment interpretation are highly sought after.  
However, while first-principles methods to predict exciton binding energies, optical transitions~\cite{rohlfing2000electron, Onida-review} and radiative lifetimes~\cite{spataru2004excitonic,palummo2015exciton,chen2018theory,jhalani2019} are well established, accurate calculations of exciton dynamics and non-radiative processes are a research frontier.\\
\indent  
The interaction between electrons and lattice vibrations (phonons) controls the dynamics of carriers and excitons. Recent advances have made \textit{ab initio} calculations of electron-phonon ($e$-ph) interactions and scattering processes widespread~\cite{bernardi2016first}, enabling studies of charge transport \cite{zhou2016ab,lee2018charge,zhou2018electron} and nonequilibrium carrier dynamics \cite{bernardi2014ab,jhalani2017ultrafast} in materials. These methods achieve quantitative accuracy, and can provide unprecedented microscopic insight into electron dynamics. In the typical workflow \cite{bernardi2016first}, one uses density functional theory (DFT) to compute the electronic band structure, and density functional perturbation theory \cite{baroni2001phonons} (DFPT) to compute phonon dispersions and the perturbation potential due to phonons. 
These quantities are combined to obtain the $e$-ph matrix elements \cite{bernardi2016first,agapito2018ab}, 
\begin{equation}
\label{Eq:eph_coupling}
\vspace{-1pt}
g_{mn,\nu}(\bfk,\bfq) = \< m \bfk+\bfq| \Delta V_{\nu \bfq } | n \bfk \>,
\end{equation} 
which represent the probability amplitude for scattering from an initial Bloch state $|n\bfk \>$ to a final state $|m \bfk+\bfq \>$, by emitting or absorbing a phonon with mode index $\nu$ and wave vector $\bfq$, due to the perturbation of the Kohn-Sham potential, 
$\Delta V_{\nu \bfq}$, induced by the phonon~\cite{bernardi2016first}.
\\
\indent
Excitons pose new challenges to this framework, since one can no longer study independently the scattering of electrons or holes with phonons when the two carriers are bound together. 
Rather, the challenge is to address exciton-phonon (ex-ph) interactions, which govern exciton dynamics over a wide temperature range, regulating photoluminescence linewidths, exciton diffusion and ultrafast dynamics~\cite{molina2017ab, mueller2018exciton, wang2018strong, wang2017raman, cannuccia2019theory, macfarlane1957fine, macfarlane1958fine, helmrich2018exciton, Bockelmann1993exciton, Remeika2013pattern,paleari2019exciton}. 
Several analytical or semi-empirical models have been proposed for ex-ph interactions~\cite{toyozawa1958theory,toyozawa1964interband,segall1968phonon,vasili2005effect,jiang2007exciton, shree2018observation, brem2018exciton, christiansen2019theory}; recent work has put forward a many-body approach but did not present numerical results~\cite{antonius2017theory}. To date, rigorous first-principles calculations of ex-ph interactions and dynamical processes are still missing. 
\\
\indent
% HERE WE SHOW
In this Letter, we derive ex-ph coupling matrix elements and relaxation times within lowest-order perturbation theory, and compute them from first principles in bulk hexagonal boron nitride (h-BN). Our results show that the ex-ph interaction can be viewed as a quantum superposition of electron and hole scattering events with phonons, weighted by the exciton wave function in the transition basis. 
Our calculations in h-BN show a dominant coupling between excitons and longitudinal optical (LO) phonons. We find ex-ph relaxation times of order 5$-$100 fs at 77 K; the relaxation times drop rapidly above the LO phonon emission threshold and become nearly temperature independent, while below the emission threshold they increase linearly with temperature. Our study provides microscopic insight into exciton thermal and dynamical processes.\\
\indent
We treat finite-momentum excitons within the \textit{ab initio} Bethe-Salpeter equation (BSE) approach, writing the BSE Hamiltonian in the transition basis (suppressing crystal momenta for now)~\cite{strinati1982dynamical,rohlfing2000electron}:
\vspace{-1pt}
\begin{equation}
\label{Eq:unper-effect-BSE-Hamil}
H_{vc,v'c'}=\<vc|H|v'c'\>=\(\e_c-\e_v\) \d_{vv'}\d_{cc'}+K_{vc,v'c'}\,\,,
\end{equation} 
where $v$ and $c$ are valence and conduction band indices, $\e_{v}$ and $\e_{c}$ the respective electron energies, and $K_{vc,v'c'}$ is the BSE kernel encoding the electron-hole interactions. 
When computing optical processes, one usually focuses on transverse excitons and removes the long-range part ($\mathbf{G}\!=\!0$ component, where $\mathbf{G}$ is a reciprocal lattice vector) of the Hartree potential from the kernel \cite{del1984macroscopic,agranovich2013crystal}. However, for ex-ph interactions, both transverse and longitudinal excitons need to be considered, so we use the full Coulomb interaction (including the $\mathbf{G}=0$ Hartree term) in the BSE kernel \footnote{private communication with Fulvio Paleari and Andrea Marini}. 
Within the Tamm-Dancoff approximation, the exciton wave function is expanded as $ |S_n\>=\sum_{vc}A^{S_n}_{vc}|vc\>$, and solving the BSE Hamiltonian in Eq.~(\ref{Eq:unper-effect-BSE-Hamil}) gives the exciton energies $E^{S_n}$ and wave function coefficients $A^{S_n}$:
\begin{equation}
\label{Eq:HAEA}
\sum_{v'c'}H_{vc,v'c'}A^{S_n}_{v'c'}=E^{S_n}A^{S_n}_{vc}.
\end{equation}
% HAMILTONIAN
\indent
To treat the ex-ph interaction, we introduce atomic displacements as a first-order perturbation to the BSE. Both the transition-basis electronic wave functions and the kernel are modified by the phonon perturbation, but the primary effect is the change in the wave functions (the change in the BSE kernel can be ignored to first order, analogous to the GW approximation~\cite{rohlfing2000electron,strinati1982dynamical}). The derivation is outlined here, and given in detail in the Supplemental Material \cite{Supplementary}. We build the BSE using the perturbed wave functions, express overlap terms using first-order perturbation theory, and quantize the resulting Hamiltonian by introducing creation and annihilation operators for phonons ($\hat{b}^\dagger$ and $\hat{b}$) and excitons ($\hat{a}^\dagger$ and $\hat{a}$).  
The exciton Hamiltonian becomes:
\begin{eqnarray}
\label{Eq:ex-Hamiltonian}
&&\tilde{H}=\sum_{n\bfQ}E_{S_n(\bfQ)}\hat{a}_{S_n(\bfQ)}^\dagger\hat{a}_{S_n(\bfQ)}
+\sum_{\v\bfq}\hbar\w_{\v\bfq}\hat{b}^\dagger_{\v\bfq}\hat{b}_{\v\bfq}
\nn\\
&&
+\sum_{nm\v,\bfQ\bfq}
\mathcal{G}_{nm\v}(\bfQ,\bfq)
\hat{a}_{S_m(\bfQ+\bfq)}^\dagger\hat{a}_{S_n(\bfQ)}
(\hat{b}_{\v\bfq}+\hat{b}^\dagger_{\v-\bfq}),
\end{eqnarray}
where $\bfQ$ is the exciton center-of-mass momentum, $S_n$ and $S_m$ label exciton states, and $\v$ is the phonon mode index. 
The second line in Eq.~(\ref{Eq:ex-Hamiltonian}) is the ex-ph interaction, with matrix elements \cite{Supplementary}
\begin{eqnarray}
\label{Eq:exph_coupling}
&&\mathcal{G}_{nm\nu}(\bfQ,\bfq)~\nn\\
&&~=\sum_{\bfk}\[
\sum_{vcc'}
A^{S_m(\bfQ+\bfq)*}
_{v\bfk,c(\bfk+\bfQ+\bfq)}
A^{S_n(\bfQ)}_{v\bfk,c'(\bfk+\bfQ)}
g_{c'c\nu}(\bfk+\bfQ,\bfq)\right.~\nn\\
&&~~~
\left.-
\sum_{cvv'}
A^{S_m(\bfQ+\bfq)*}_{v(\bfk-\bfq),c(\bfk+\bfQ)}
A^{S_n(\bfQ)}_{v'\bfk,c(\bfk+\bfQ)}
g_{vv'\nu}(\bfk-\bfq,\bfq)
\]
\end{eqnarray}
that quantify the probability amplitude for scattering from an initial exciton state $|S_n\>$ with momentum $\bfQ$ to a final state $|S_m\>$ with momentum $\bfQ+\bfq$ due to absorption or emission of a phonon with mode index $\nu$ and wave vector $\bfq$; the $e$-ph coupling matrix elements, $g$, are defined above in Eq.~(\ref{Eq:eph_coupling}).
This ex-ph coupling, which is pictorially shown in Fig.~\ref{Fig:scattering}, is a quantum superposition of electron- and hole-phonon scattering processes, weighted by the exciton wave functions of the initial and final states. 

\begin{figure}[t]
\includegraphics[scale=0.235]{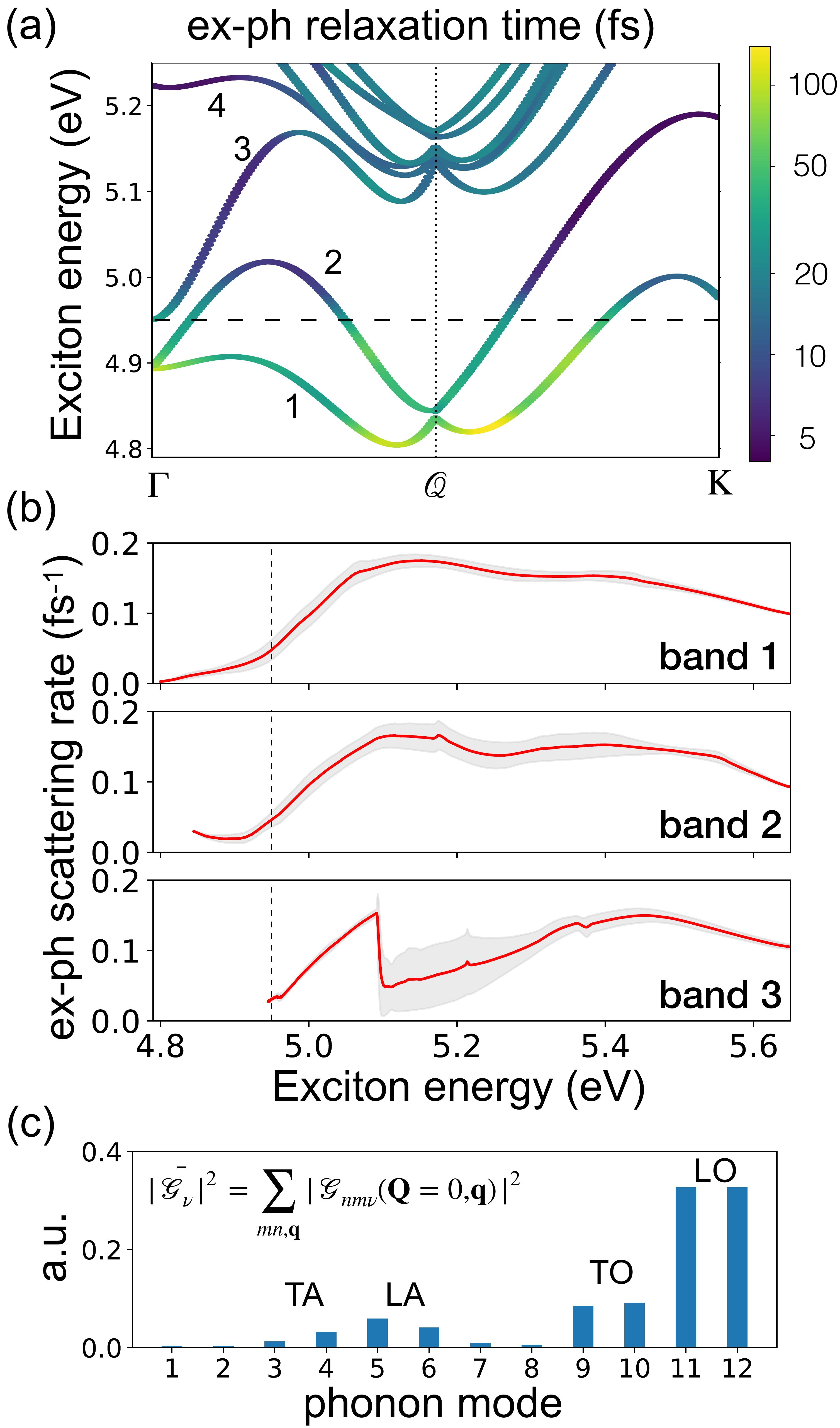}
\caption{(a) Exciton band structure for the four lowest-energy exciton bands, together with a log-scale color map of ex-ph relaxation times at 77 K. Note the drop in the relaxation times above the LO emission threshold at 4.95 eV, which is shown with a dashed line. (b) Average ex-ph scattering rates as a function of exciton energy, up to 5.7 eV for three chosen exciton states. The shaded region gives the standard deviation of the momentum-dependent scattering rate at each energy. 
(c) The squared ex-ph coupling strength (in arbitrary units) for the 12 phonon modes of h-BN.}
\label{Fig:E-rate}
\end{figure}

\begin{figure}[!t]
\includegraphics[scale=0.295]{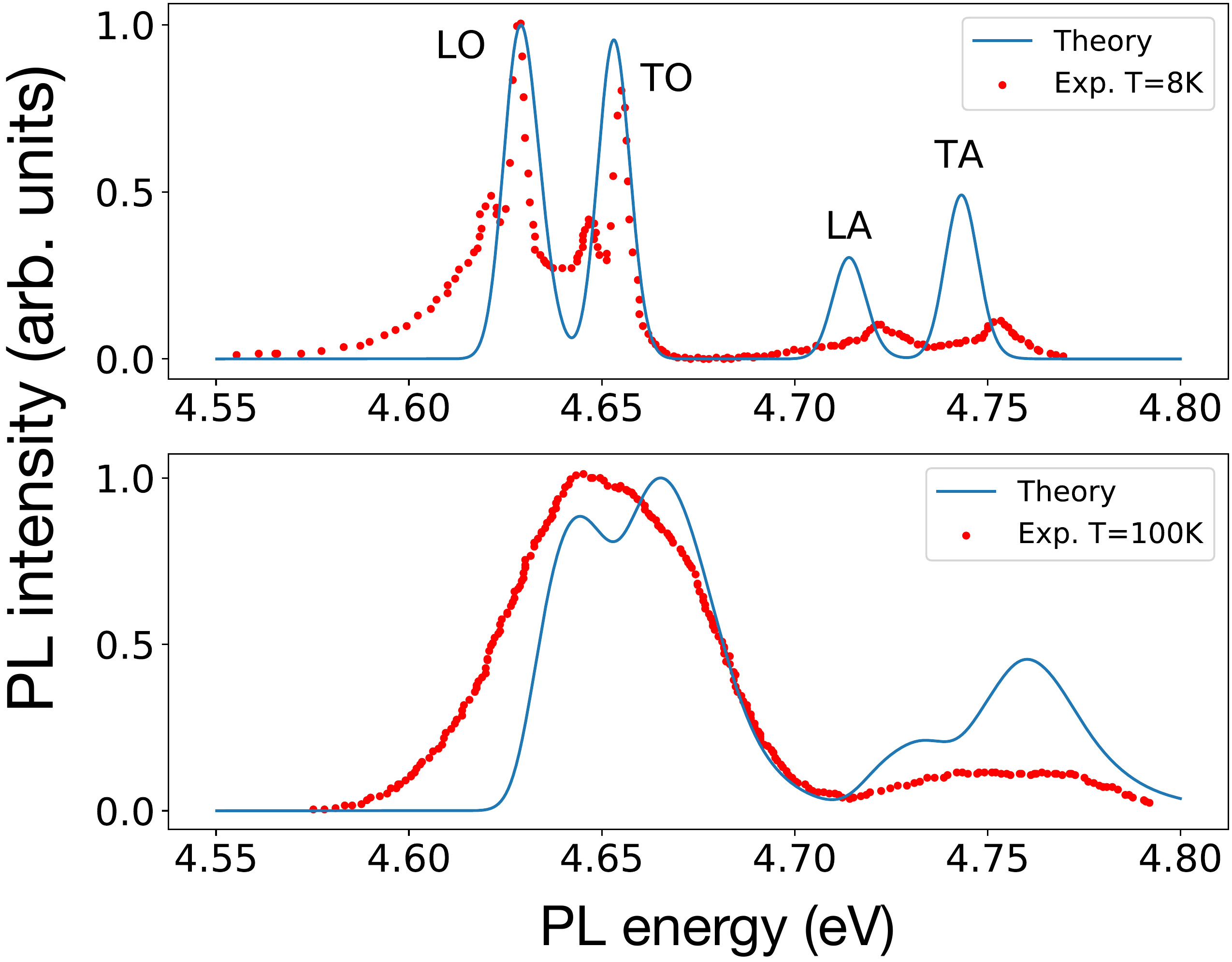}
\caption{Comparison between the computed phonon-assisted PL intensity (blue curve) and experimental data from Ref.~\cite{PL-measurement2017} (red dots). 
The calculated spectra were shifted and normalized to match the first peak of the measured spectra.}
\label{Fig:PL}
\end{figure}

%
% EX-PH SCATTERING RATE
%
\indent
Analogous to the case of $e$-ph interactions, we compute the ex-ph scattering rate at temperature $T$, $\Gamma^{\rm ex-ph}_{n\bfQ}(T)$, and its inverse, the ex-ph relaxation time, $\tau_{n \bfQ}(T) =1/ \Gamma_{n\bfQ}(T)$, obtaining \cite{Supplementary}:
\begin{eqnarray}
\label{Eq:relaxation time}
&&\Gamma^{\rm ex-ph}_{n\bfQ}(T)=\frac{2\pi}{\hbar} \frac{1}{\mathcal{N}_\bfq}
\sum_{m\nu\bfq}|\mathcal{G}_{nm\nu}(\bfQ,\bfq)|^2\nn\\
&&~\times\[(N_{\nu\bfq}+1+F_{m \bfQ+\bfq})\times\d (E_{n\bfQ}-E'_{m\bfQ+\bfq}-\hbar\w_{\nu \bfq})\right.\nn\\
&&~~~~
\left.+(N_{\nu\bfq}-F_{m \bfQ+\bfq})\times
\d (E_{n\bfQ}-E'_{m\bfQ+\bfq}+\hbar\w_{\nu \bfq})
\],
\end{eqnarray}  
where $N$ are phonon and $F$ exciton (bosonic) occupation factors, and $\mathcal{N}_\bfq$ is the number of $\bfq$-points~\footnote{Although obtained here with Rayleigh-Schrodinger perturbation theory, Eqs.~(\ref{Eq:exph_coupling})-(\ref{Eq:relaxation time}) agree with the results from the many-body treatment in Ref.~\cite{antonius2017theory}.}. In our approach, the temperature dependence of the relaxation times is due to the phonon and exciton occupation factors, while the exciton wave functions and energies are computed with the BSE on a fixed atomic structure at zero temperature.
\\
\indent 
The numerical calculations on h-BN are carried out within the local density approximation of DFT using the {\sc{Quantum Espresso}} code \cite{giannozzi2009quantum}. 
We use norm-conserving pseudopotentials~\cite{perdew1981self,troullier1991efficient} and a 60 Ry kinetic energy cutoff to compute the electronic structure (with DFT) and lattice vibrations (with DFPT). The $e$-ph calculations are carried out with the {\sc{Perturbo}} code~\cite{perturbo}, while GW and finite-momentum BSE calculations are carried out with the {\sc{Yambo}} code~\cite{sangalli2019many}. The same $24 \times 24 \times 4$ Brillouin zone grid is used for $\bfk$-points (for electrons), $\bfq$-points (for phonons) and $\bfQ$-points (for excitons). 
The ex-ph matrix elements are computed without interpolation or symmetry. For the ex-ph scattering rates, we use linear interpolation to obtain the matrix elements and exciton energies on a $120\times 120 \times 20$ Brillouin zone grid. Additional numerical details, phonon dispersion, and convergence analysis are also provided~\cite{Supplementary}. %in Eq.~(\ref{Eq:relaxation time}) of the ex-ph scattering rates 
\\  
\indent
Figure \ref{Fig:E-rate}(a) shows the exciton band structure, defined here as the exciton energy versus momentum dispersion curves, along a high-symmetry line for the lowest 8 exciton bands (numbered in order of increasing energy). 
Overlaid to the exciton band structures are the ex-ph relaxation times at $77$ K. Our exciton band structure agrees well with previous results~\cite{sponza2018exciton,sponza2018direct}, apart from a small rigid energy shift; the global minimum is located close to a point called here $\mathcal{Q}$, the halfway point between $\Gamma$ and K, which corresponds to the excitation across the indirect electronic band gap of h-BN. Note also that in our calculation the degeneracy between the 3rd and 4th exciton bands at $\Gamma$ is lifted due to the inclusion, different from Refs.  \cite{sponza2018exciton,sponza2018direct}, of the $\mathbf{G}=0$ Hartree term in the BSE kernel, which splits transverse and longitudinal excitons. 
\\
\indent
Our computed ex-ph relaxation times are of order 5$-$100 fs over a wide temperature range up to 300~K, corroborating the widely used assumption that excitons thermalize rapidly before recombining. 
The relaxation times are strongly energy dependent. At 77 K, they are of order 100 fs near the exciton energy minima, and drop rapidly to $\sim$15 fs above the threshold for LO phonon emission, 
located 160 meV above the exciton energy minima [at exciton energy of 4.95 eV; see Fig.~\ref{Fig:E-rate}(a)]. 
Analysis of the ex-ph coupling strength [Fig.~\ref{Fig:E-rate}(c)] and scattering rate due to each individual phonon mode reveals that above this threshold the strongest scattering channel is the emission of an LO phonon with average energy of 160 meV \cite{Supplementary}. Since the exciton energy minimum is at 4.8 eV, only excitons with energy greater than 4.95 eV can emit an LO phonon and scatter to a final exciton state, which explains the much shorter relaxation times above the LO emission threshold. This trend is analogous to the $e$-ph scattering rates in polar semiconductors (e.g., GaAs), where electrons couple strongly with LO phonons and the relaxation time drops rapidly above the LO emission threshold~\cite{zhou2016ab}.
\\
\indent
% COUPLING STRENGTH
\begin{figure}[b]
\centering
\includegraphics[scale=0.26]{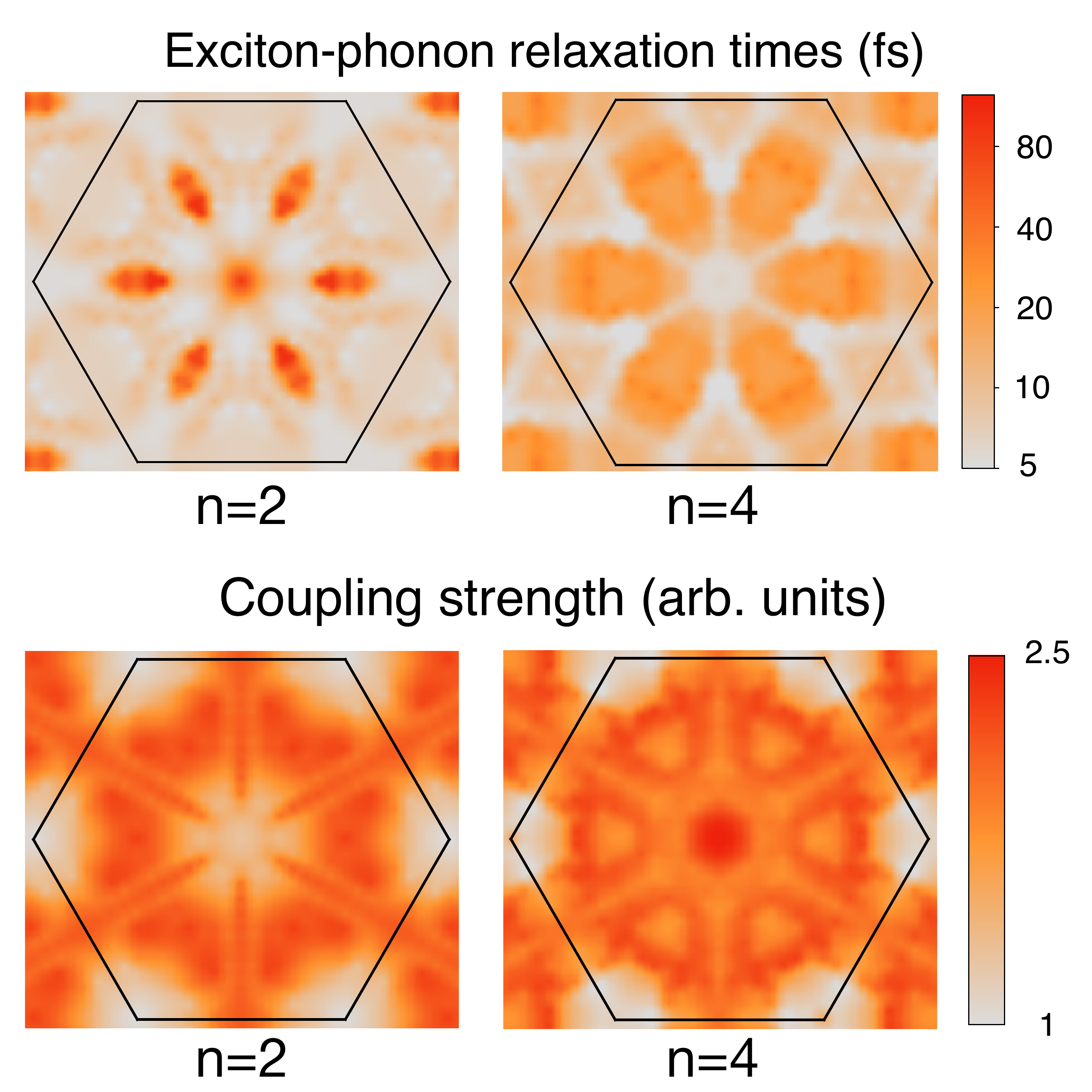}  
\caption{Exciton-phonon relaxation times at 77 K, shown as a function of exciton momentum in the Brillouin zone plane parallel to the h-BN layers. 
The lower panel shows the average coupling strength $\bar{\mathcal{G}}_n(\bfQ)$ in arbitrary units, where red color indicates stronger coupling. In both panels, the color maps are given on a log-scale.}
\label{Fig:tau-momentum}
\end{figure}
Figure \ref{Fig:E-rate}(b) shows the ex-ph scattering rates as a function of exciton energy, averaged over exciton momentum, for exciton bands 1$-$3. For the lowest-energy excitons in band 1, the scattering rate increases monotonically with energy between 4.8$-$5.05 eV, with a change of slope at 4.95 eV due to the onset of LO phonon emission. We find an LO phonon emission time of $\sim$15 fs for excitons at 77~K, a value comparable to LO phonon emission times for electrons in polar semiconductors~\cite{zhou2016ab}. 
Compared to excitons in band 1, the scattering rate is higher at low energy for excitons in band 2, which can emit phonons with a range of energies and transition to band 1. 
Excitons in band 3 exhibit a drop in the scattering rate at 5.1 eV due to the energy minima near $\mathcal{Q}$ with significant energy gaps from the two lower bands.\\
\indent
% TEMPERATURE DEPENDENCE
The momentum dependence of the relaxation times is controlled by the exciton band structure, which provides the phase space for scattering, and by the ex-ph matrix elements. 
Figure~\ref{Fig:tau-momentum} analyzes the exciton relaxation times (for two specific bands, 2 and 4) as a function of exciton momentum in the Brillouin zone, 
together with the average ex-ph coupling strength, defined as $\bar{\mathcal{G}}_n(\bfQ)=\sum_{m\nu\bfq} |\mathcal{G}_{nm\nu} (\bfQ,\bfq)|^2$. Both the ex-ph coupling and relaxation times exhibit the six-fold symmetry of h-BN. 
As a general trend, we find that larger coupling strengths are associated with shorter relaxation times, consistent with Eq. (\ref{Eq:relaxation time}). The relaxation times are maximal near the exciton local energy minima at $\mathcal{Q}$ 
(and also at $\Gamma$ for band 2), where the anisotropic exciton dispersion gives rise to ellipsoid-shaped regions in momentum space with longer relaxation times.
\\
\indent
\begin{figure}[!b]
\includegraphics[scale=0.29]{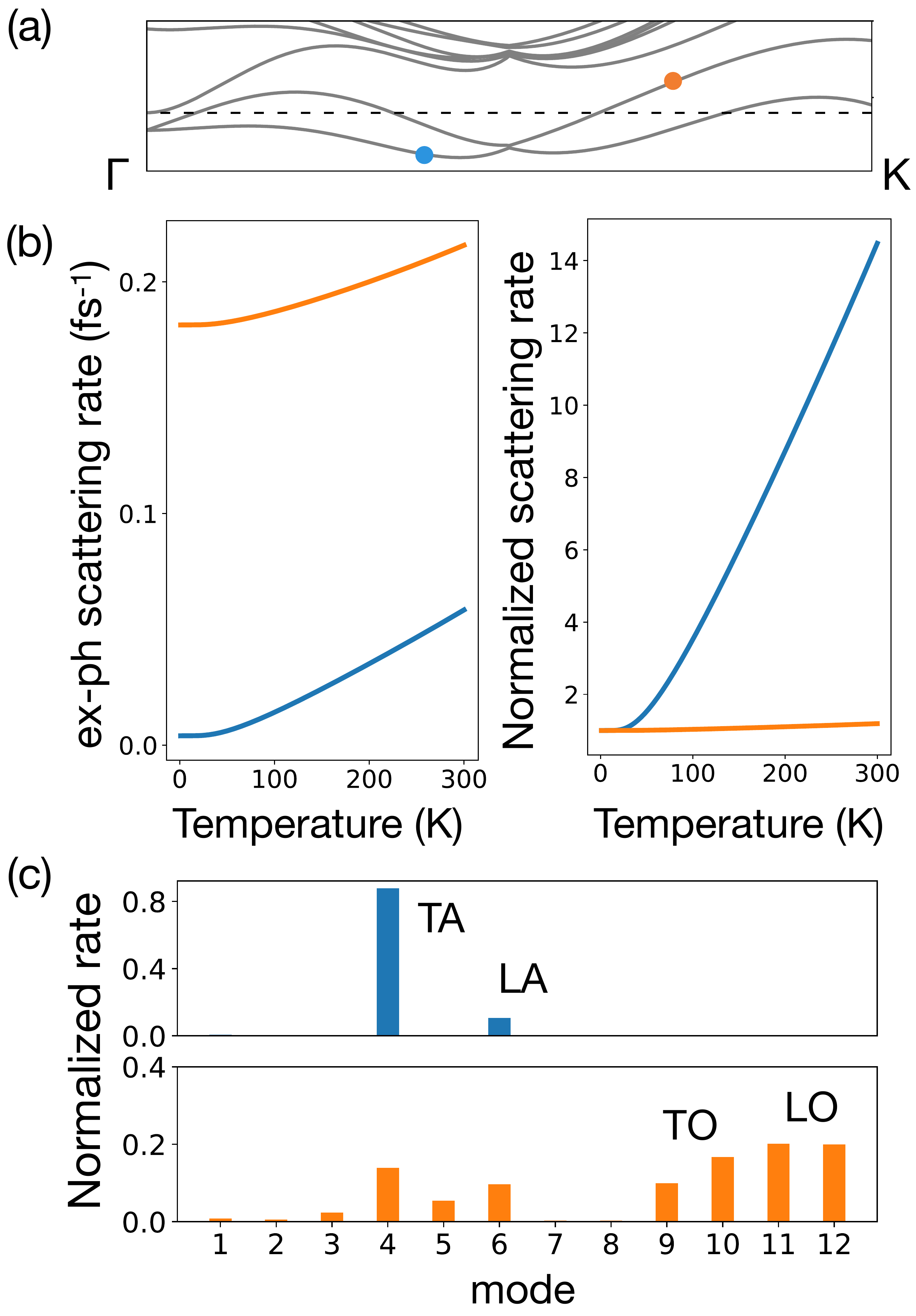}
\caption{Ex-ph scattering rate as a function of temperature for an exciton state below (blue) and above (orange) the LO-emission threshold. (a) The two states selected for the analysis are shown in the exciton band structure. (b) The absolute value of the scattering rates (left) and the same quantities normalized by the scattering rate at 1 K for each state (right) to emphasize the stronger temperature dependence for the state below the LO emission threshold. (c) Mode-resolved contribution to the scattering rate at 1 K, normalized by the total rate for each state.}
\label{Fig:tau-T}
\end{figure}
Understanding how temperature affects exciton dynamics is crucial in experiments. Figure~\ref{Fig:tau-T} compares the temperature dependence of the ex-ph scattering rates for two exciton states, one above and one below the LO-phonon emission threshold. The scattering rate increases monotonically from 1 to 300 K for both states, but with rather different trends. For the state below the LO emission threshold, the scattering rate increases by over an order of magnitude between 1$-$300 K, while the increase for the state above the LO emission threshold is much smaller, only about 10 percent over the same temperature range. We find similar trends when inspecting other states in these two energy windows.
Analysis of the contributions to exciton scattering from the different phonon modes [see Fig.~\ref{Fig:tau-T}(c)] reveals that excitons below the LO emission threshold and close to the energy minima mainly scatter by absorbing low-energy acoustic phonons, which explains the strong temperature dependence. On the other hand, at energies above the LO emission threshold, scattering is dominated by LO phonon emission, a weakly temperature dependent process with rate proportional to $N$+1 [see Eq.~(\ref{Eq:relaxation time})].\\
\indent
% PL ANALYSIS
Using the ex-ph matrix elements, we can also compute the phonon-assisted photoluminescence (PL) spectrum in h-BN~\cite{Supplementary}, 
obtaining results in agreement with PL experiments between 8$-$100~K (see Fig.~\ref{Fig:PL}).  
At low temperature of 8 K, our computed PL exhibits all four peaks seen in experiment, which correspond to LO, TO, LA and TA phonon-assisted PL. 
We find dominant LO and TO peaks due to the strong ex-ph coupling of these phonon modes. 
At 100~K, the PL peak linewidths accurately match the experimental data.  
Yet, the computed acoustic peaks are too intense, and the relative LO and TO peak intensities at 8~K (but not at 100~K) are sensitive to the broadening used in the calculations. 
Additional work is needed to fully converge these fine features of the PL spectra. 
\\
\indent

% SUMMARY
In summary, we derived an approach for computing the interaction between excitons and phonons and the associated matrix elements and relaxation times. 
Our calculations in h-BN reveal the dominant ex-ph coupling with the LO mode, 
identify the threshold for LO phonon emission and the associated $\sim$15 fs LO emission time, and unravel the momentum, energy and temperature dependence of ex-ph scattering processes. 
Our approach paves the way to quantitative studies of exciton transport and ultrafast dynamics in materials with strongly \mbox{bound excitons}. We plan to implement a numerical scheme for real-time exciton dynamics using the ex-ph interactions derived in this work.\\ 
\indent

\vspace{-10pt}
We thank F. Paleari and A. Marini for a critical reading of the manuscript. This work was supported by the Department of Energy under Grant No. DE-SC0019166, which provided for theory and method development. 
The code development was supported by the National Science Foundation under Grant No. ACI-1642443. 
D.S. acknowledges funding from MIUR PRIN Grant No. 20173B72NB and by the European Union's Horizon 2020 research and innovation program (Grants No. 824143 and No. 654360). This research used resources of the National Energy Research Scientific Computing Center, a DOE Office of Science User Facility supported by the Office of Science of the US Department of Energy under Contract No. DE-AC02-05CH11231.
\bibliographystyle{apsrev4-1}
%\bibliography{exph_ref}
%merlin.mbs apsrev4-1.bst 2010-07-25 4.21a (PWD, AO, DPC) hacked
%Control: key (0)
%Control: author (72) initials jnrlst
%Control: editor formatted (1) identically to author
%Control: production of article title (-1) disabled
%Control: page (0) single
%Control: year (1) truncated
%Control: production of eprint (0) enabled
%

\end{document}